\documentclass[12pt]{iopart}

\usepackage{graphicx}
\usepackage{color}

\newcommand{\avg}[1]{\left\langle #1 \right\rangle}

\newcommand{\abs}[1]{\left| #1 \right|}
\newcommand{\eq}[1]{Eq.~(\ref{#1})}
\newcommand{\twoEq}[2]{Eqs.~(\ref{#1}) and (\ref{#2})}

\newcommand{\fig}[1]{Fig.~\ref{#1}}
\newcommand{\secref}[1]{Sec.~\ref{#1}}

\newcommand{\bra}[1]{ \left\langle #1 \right |}
\newcommand{\ket}[1]{ \left |#1 \right\rangle}

\def \TT{{\mathcal T}}

\def \half{{\frac{1}{2}}}

\def \figwidth {0.6\textwidth}

\def \wide {0.9\textwidth}

\def \nth {\bar n_{\rm th}}

\def \sz {\hat \sigma_z}
\def \sx {\hat \sigma_x}
\def \sy {\hat \sigma_y}
\def \nimp {\bar n_{\rm min}}
\def \nadd {\bar n_{\rm ba}}
\def \nmeas{\bar n_{\rm meas}}
\def \nssr{\bar n_{\rm min,1}}
\def \Nopt {N_{\rm opt}}

\def \Qeff {Q_{\rm eff}}
\def \ttot {{t_{\rm tot}}}
\def \sens {\xi}
\def \Gphi {\Gamma_\phi}

\begin{document}

\title{Measuring mechanical motion with a single spin}

\author{S D Bennett$^1$, S Kolkowitz$^1$, 
Q P Unterreithmeier$^1$,
P Rabl$^2$, \\
A C Bleszynski Jayich$^3$, J G E Harris$^4$ 
and M D Lukin$^1$}
\address{$^1$ Department of Physics, Harvard University, Cambridge, MA 02138, USA}
\address{$^2$ Institute for Quantum Optics and Quantum Information of the Austrian Academy of Science, 6020 Innsbruck, Austria}
\address{$^3$ Department of Physics, University of California Santa Barbara, 
Santa Barbara, \\ CA 93106, USA}
\address{$^4$ Departments of Physics and Applied Physics, Yale University, 
New Haven, \\ CT 06520, USA}

\ead{bennett@physics.harvard.edu}

\date{\today}

\begin{abstract}
We study theoretically the measurement of
a mechanical oscillator
using a single two level system as a detector.
In a recent experiment, we
used a single electronic spin associated with a
nitrogen vacancy center in diamond to probe the
thermal motion of a magnetized
cantilever at room temperature 
\{Kolkowitz {\it et al.}, {\it Science} {\bf 335}, 1603 (2012)\}.
Here, we present a detailed analysis
of the 
sensitivity limits of this technique,
as well as the 
possibility to measure the zero point motion
of the oscillator.
Further, we discuss the issue of 
measurement backaction
in sequential measurements and find
that although backaction heating can occur,
it does not prohibit the detection
of zero point motion.
Throughout the paper we focus on 
the experimental implementation of
a nitrogen vacancy center coupled
to a magnetic cantilever; however,
our results are applicable 
to a wide class of spin-oscillator
systems.
Implications for preparation of nonclassical
states of a mechanical oscillator are also
discussed.

\end{abstract}
\maketitle

\section{Introduction}

Recent interest in mechanical oscillators
coupled to quantum systems is motivated
by quantum device applications
and by the goal of observing quantum 
behavior of macroscopic mechanical objects.
The past decade has seen
rapid progress 
studying
mechanical oscillators coupled to
quantum two-level systems such as
superconducting qubits 
\cite{LaHaye:2009ja,OConnell:2010br,Armour:2002kta},
and single electronic spins
\cite{Rugar:2004cr},
and theoretical work has explored 
strong mechanical coupling 
to collective
atomic spins \cite{Treutlein:2007km,Steinke:2011ig}.
Recently, it was proposed that a mechanical
oscillator could be strongly coupled to
an individual spin qubit
\cite{Rabl:2009fz,Palyi:2011vla}.
Experiments based on
single spins coupled to mechanical systems have
demonstrated scanning magnetometry
\cite{Balasubramanian:2008ga},
mechanical spin control \cite{Hong:2012wr}, 
and detection of mechanical motion 
\cite{Arcizet:2011cg,Kolkowitz:2012iw}.
In parallel,
pulsed spin control techniques have attracted
renewed interest for
decoupling a spin from low
frequency noise in its environment,
extending its coherence \cite{deLange:2010ga},
while also enhancing the sensitivity of the 
spin for magnetometry
\cite{Maze:2008ws, Lange:2011jo, Naydenov:2011eo, Laraoui:jk}.

In this paper we
consider pulsed single spin 
measurements applied to the detection of
mechanical motion at the single phonon level.
We extend the analysis presented in our
recent work \cite{Kolkowitz:2012iw},
providing a detailed theoretical
framework and a discussion
of measurement backaction. 
The central concept of our measurement
approach is to apply a sequence of
control pulses to the spin, synchronizing its
dynamics with the period
of a magnetized cantilever, thereby enhancing its
sensitivity to the motion.
By measuring the variance of the accumulated
phase imprinted on the spin by the oscillator
during a measurement, we directly probe
the average phonon number, 
despite the fact that the oscillator position 
is linearly coupled to the transition frequency of the spin.
We derive the conditions for observing a single
phonon using
the spin as a detector, and find that these conditions
coincide with that of large effective cooperativity, 
sufficient to perform a two-spin gate
mediated by mechanical motion \cite{Rabl:2010gza}.
Further, we consider the
backaction arising from sequential measurements
and show that this does not prohibit single
phonon resolution.
Throughout the paper, 
we focus on the specific spin-oscillator system
of a magnetized cantilever
coupled to the electronic spin associated
with a nitrogen-vacancy (NV) center in diamond.
For realistic experimental parameters
we find that this system can
reach the regime of large cooperative
spin-phonon coupling, and the spin
may be used to
measure and manipulate 
mechanical motion at the quantum level.

We begin in \secref{sec:dd} by introducing
the coupled system
and spin control sequences,
and calculate the
signal due to thermal and driven
motion of the oscillator.
Then in \secref{sec:sensitivity}
we derive the optimal phonon number 
sensitivity, and show the relation between
strong cooperativity and single phonon resolution.
Finally, in \secref{sec:zpm}
we consider the limit of zero temperature
and calculate
the signal due to zero point motion,
including a discussion of
backaction heating for sequential measurements.

\section{Coherent sensing of mechanical motion}
\label{sec:dd}

\subsection{Model}

\begin{figure}[tb]
\begin{centering}
	\includegraphics[height=4cm]{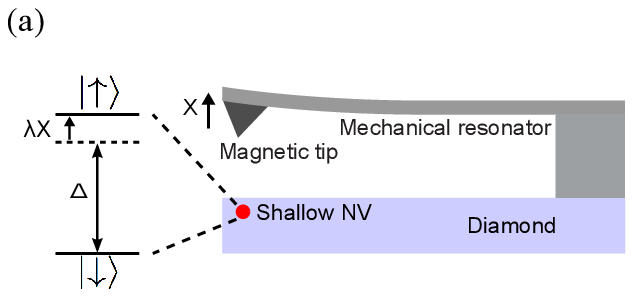}
	\hspace{1cm}
        	\includegraphics[height=4cm]{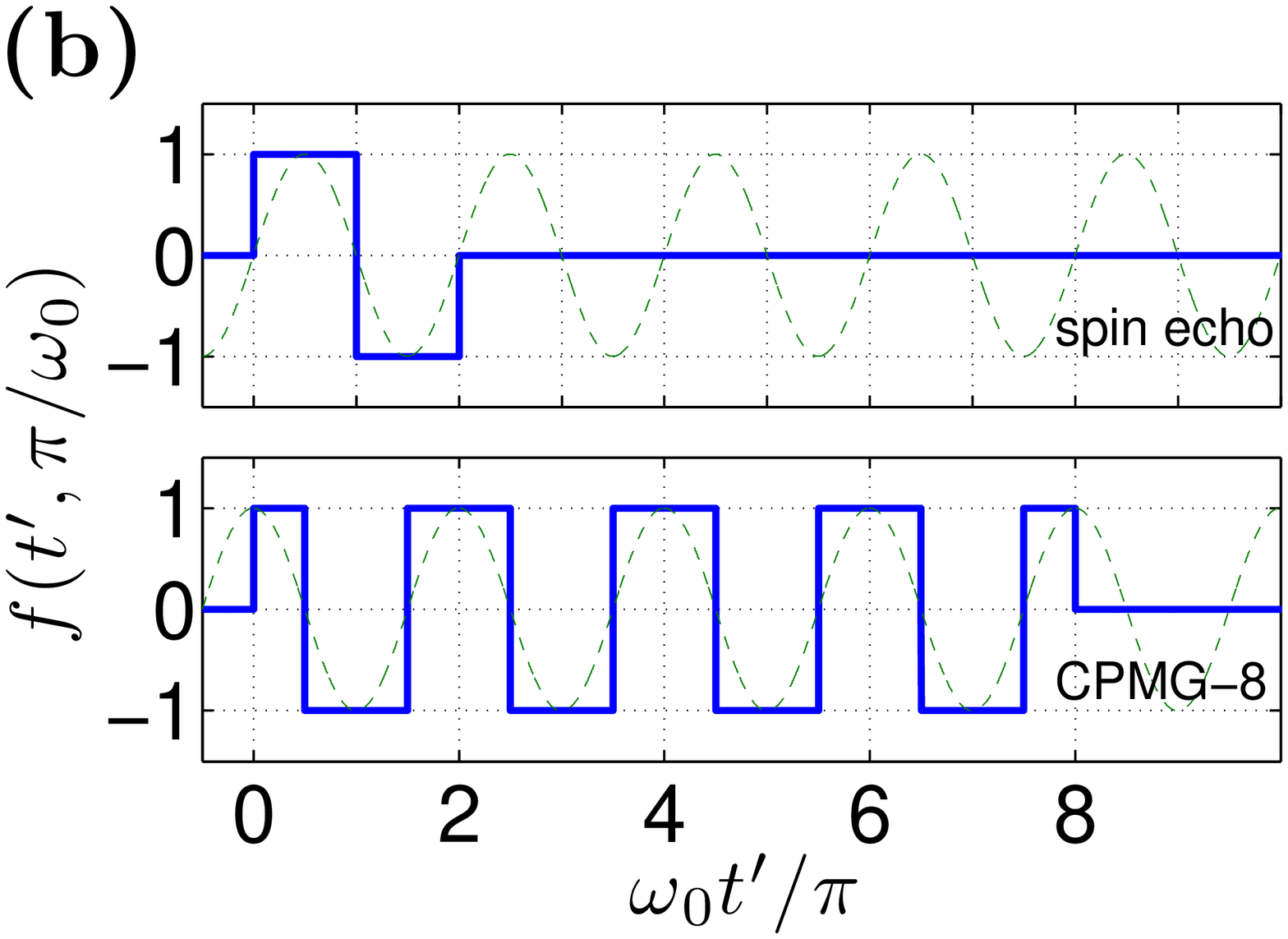}
\caption{{\bf(a)} Schematic of the setup.
A single spin can be used to measure mechanical
motion via magnetic coupling.
{\bf(b)} Toggling sign of the interaction describing
$\pi$ pulses flipping the spin.
Each sequence begins and ends with $\pi/2$ pulses,
and $\pi$ pulses
flip the sign of the interaction at regular intervals
of time $\tau$.
Thin dashed line shows oscillator position, 
which is synchronized with pulse
sequence for $\tau = \pi/\omega_0$ as shown.
The total sequence time 
is $t = 2\tau$ for spin echo
and $t = N\tau$ for CPMG.
}
\label{fig:cartoon}
\end{centering}
\end{figure}

We consider the setup shown 
schematically in \fig{fig:cartoon}, in which
a magnetized cantilever is coupled to the electronic
spin of a single NV center.
The magnetic tip generates a field gradient at the location of the NV,
and as a result its motion
modulates the magnetic field seen by the spin
causing Zeeman shifts of its precession frequency.
To lowest order in small cantilever motion, the precession frequency
depends linearly on
the position of the tip and is
described by the Hamiltonian
($\hbar = 1$)
\begin{equation}
\label{eq:H}
     \hat H = \frac{\Delta}{2}  \sz +
    \frac{\lambda}{2}  \left(  \hat a +  \hat a^\dagger \right)
      \sz
    +  \hat H_{\rm osc},
\end{equation}
where $\sz$
is the Pauli operator
of the spin
and $\hat a$ is the annihilation operator
of the oscillator.
For a spin
associated with an NV center in diamond,
we take
$\ket{\uparrow} = \ket{m_s = 1}$
and $\ket{\downarrow} = \ket{m_s = 0}$
in the spin-1 ground state of the NV center,
and safely ignore the $\ket{m_s = -1}$ state
assuming it is far detuned
by an applied dc magnetic field.
$\Delta$ is the detuning
of the microwave pulses used for
spin manipulation, which plays no role
in what follows and we take
$\Delta = 0$ throughout the paper.
The spin-oscillator coupling strength is
$\lambda = g_e \mu_B G_m x_0/\hbar$,
where $g_e\approx 2$ is the Land\'e $g$-factor, $\mu_B$
is the Bohr magneton, $G_m$ is the magnetic
field gradient along the NV axis, and
$x_0 = \sqrt{\hbar/ 2 m \omega_0}$ is
the zero point motion of the cantilever mode of mass $m$
and frequency $\omega_0$
(we included $\hbar$ in the
definitions of $\lambda$ and $x_0$ for clarity).
The damped, driven oscillator is described by
\begin{equation}
\label{eq:Hosc}
     \hat H_{\rm osc} = \omega_0  \hat a^\dagger  \hat a
    + \hat H_\gamma + \hat H_{\rm dr},
\end{equation}
where
\begin{equation}
    \hat H_\gamma =
    \sum_k g_k
    \left(  \hat a +  \hat a^\dagger \right)
    \left(  \hat b_k + \hat b^\dagger_k \right)
    + \sum_k \omega_k  \hat b^\dagger_k  \hat b_k
\end{equation}
describes dissipative coupling to a
bath of oscillators  $\hat b_k$,
characterized by damping rate $\gamma$
and temperature $T$.
Finally,
$\hat H_{\rm dr}$ describes a coherent
oscillator drive which we consider briefly in
\secref{sec:driven}.
Note that in \eq{eq:H} we have temporarily omitted
intrinsic spin decoherence due to the environment;
we will include this explicitly in \secref{sec:sensitivity}.
% ======================== figure =================
% ============================================

\subsection{Spin echo and multipulse sequences}
\label{sec:echo}

The motion of the oscillator imprints a
phase on the spin as it evolves under \eq{eq:H}, 
which can be detected using
spin echo \cite{Armour:2002kta,Armour:2008kg},
or more generally a multiple pulse measurement.
Throughout the paper we focus 
on Carr-Purcell-Meiboom-Gill (CPMG) type
pulse sequences, consisting of equally spaced
$\pi$ pulses at intervals of time $\tau$,
as depicted in \fig{fig:cartoon}.
After initialization in $\ket{\uparrow}$,
a $\pi/2$ pulse prepares the spin in an
eigenstate of $\sx$,
$\ket{\psi_0} = \half (\ket{\uparrow} + \ket{\downarrow})$
with $\bra{\psi_0}\sx\ket{\psi_0} = 1$.
The spin is then allowed to interact
with the oscillator for time $t$,
accumulating a phase, and during which time
we apply a
sequence of $\pi$ pulses 
which effectively
reverse the direction of spin precession.
At the end of the sequence, 
a final $\pi/2$ pulse converts the accumulated
phase into a population in $\ket{\uparrow}$
which is then read out.
By applying both initial and final $\pi/2$ rotations 
about the same axis,
we measure the probability
to find the spin in its initial state $\ket{\psi_0}$
at the end of the sequence, given by
\begin{equation}
\label{eq:S}
    P(t) = \frac{1}{2}
    \big( 1 +  \avg{\sx(t)} \big),
\end{equation}
where angle brackets denote
the average over spin and oscillator degrees of freedom.
Our choice to measure $\sx$ probes
the accumulated phase variance;
this is crucial for our purpose because the average
phase imprinted by an undriven fluctuating
oscillator is zero. 
In contrast, by applying the first and final $\pi/2$ pulses about
orthogonal axes
one would instead measure $\sy$, which probes
the average accumulated phase.

The sensitivity of the spin to mechanical motion is determined
by the impact of the oscillator on the spin coherence $\avg{\sx(t)}$.
The key to maximizing this impact is to  
synchronize the spin evolution with the mechanical period
using a CPMG sequence of $\pi$ pulses,
increasing the accumulated phase variance and improving
the sensitivity as discussed in the context of ac 
magnetometry \cite{Taylor:2008cp}.
Choosing $\tau=\pi/\omega_0$ between the $\pi$
pulses, 
we flip the spin every half-period of the oscillator
and maximize the accumulated phase variance.
At the same time, these pulse sequences 
decouple the spin from
low-frequency magnetic noise of the environment, 
extending the spin coherence
time $T_2$ \cite{Lange:2011jo,Naydenov:2011eo}.
We describe the effects of the
applied $\pi$ pulses
using a
function $f(t,\tau)$, which flips
the sign of the spin-oscillator interaction
at regular intervals of time $\tau$
as illustrated in \fig{fig:cartoon}.
In this toggling frame,
the interaction Hamiltonian is
\begin{equation}
\label{eq:Hint}
     \hat H_{\rm int} (t) = \frac{\lambda}{2}
      \sz
     \hat X(t) f(t,\tau),
\end{equation}
where $\hat X =  \hat a +  \hat a^\dagger$ and
$ \hat X(t) = e^{i   \hat H_{\rm osc} t}  \hat X  e^{-i   \hat H_{\rm osc} t}$.
We calculate the spin coherence,
$\avg{\sx(t)} = \avg{ U^\dagger(t) \sx U(t)}$,
where the evolution operator is
$\hat U(t) = \TT e^{-\frac{i \lambda}{2} \sz \int_0^t dt' \hat X(t') f(t',\tau)}$ and
$\TT$ denotes time ordering.
Since the interaction is proportional to $\sz$, it leads
to pure dephasing and we
obtain \cite{Makhlin:2004km}
\begin{equation}
\label{eq:D}
    \avg{\sx(t)}
    =
    \avg{
    \tilde \TT  e^{-i \hat \phi/2}
    \TT e^{-i \hat \phi/2}}_{\rm osc},
\end{equation}
where 
we used $\avg{\sx(0)} = 1$,
the average $\avg{\cdot}_{\rm osc}$ is
over oscillator degrees of freedom,
$\tilde \TT$ denotes anti-time ordering, and the accumulated phase operator is
\begin{equation}
\label{eq:phase}
     \hat \phi = \lambda \int_0^t  dt'   \hat X(t') f(t',\tau).
\end{equation}
The spin coherence in
\eq{eq:D} can be calculated
using a cumulant expansion, which
is vastly simplified by noting that
the full Hamiltonian in \eq{eq:H},
including the oscillator drive and ohmic dissipation,
is quadratic in
 $\hat X$.
As a result, the second
cumulant---which
in general corresponds to a Gaussian
approximation---in the present case constitutes the exact result.
We use this below to calculate
the coherence
for both thermal and driven motion.

Another consequence of the fact that $\hat H$ is quadratic
in $\hat X$ is that the effect of the
pulse sequence
is completely characterized
by its associated filter function \cite{Taylor:2008cp,deSousa:2009fp},
$F(\omega \tau) = \frac{\omega^2}{2} \big| \tilde f(\omega) \big|^2$
with $\tilde f(\omega) = \int dt e^{i\omega t} f(t,\tau)$.
The filter function describes how two-time 
position correlations $\avg{\hat X(t) \hat X(t')}$ of the oscillator
affect the spin coherence
in the second cumulant in the expansion
of \eq{eq:D}.
For the pulse sequences
illustrated in \fig{fig:cartoon}, 
the corresponding filter functions are
\begin{equation}
    F(\omega\tau) =
    \cases{
        8 \sin^4(\omega\tau/2) &\quad{\rm spin echo} \\
        2 \sin^2\left(N\omega\tau / 2 \right)
            \left[ 1-\sec\left( \omega\tau / 2 \right)\right]^2
&\quad{\rm CPMG}
    }
\end{equation}
Note that phase-alternated versions of
CPMG, such as XY4, which vary the
axis of $\pi$ pulse rotation
in order to
mitigate pulse
errors,
are also described by the above model
in the limit of ideal pulses.

\subsection{Thermal motion }
\label{sec:mech}

As discussed above, the spin coherence 
in \eq{eq:D} is given exactly
by its second order
cumulant expansion.
Since the total sequence time is $t = N \tau$,
the coherence depends only on the time
$\tau$ between $\pi$ pulses,
\begin{equation}
\label{eq:Dtherm}
	\avg{\sx(t = N\tau)} = e^{-\chi_N(\tau)},
\end{equation}
where
\begin{equation}
	\chi_N(\tau) =
	\lambda^2 \int \frac{d\omega}{2\pi}
	\frac{	F(\omega \tau)}{\omega^2}
	\bar S_X(\omega),
\label{eq:chi}
\end{equation}
and $\bar S_X(\omega) = \int dt e^{i\omega t} 
\half \langle\{   \hat X(t),   \hat X(0)\}\rangle$ 
is the symmetrized  noise
spectrum of $\hat X$.
For the damped thermal oscillator
described by $\hat H_{\rm osc}$ in the
abscence of a drive,
the symmetrized spectrum is ($k_B = 1$)
\begin{equation}
\label{eq:SX}
	\bar S_X(\omega) 
	= \frac{ 2 \omega_0 \gamma \omega \coth(\omega/2T) }
	{(\omega^2-\omega_0^2)^2 + \gamma^2 \omega^2},
\end{equation}
where $\gamma = \omega_0/Q$ is the mechanical
damping rate
due to coupling to the ohmic environment at temperature $T$.
% ======================== figure =================
\begin{figure}[htb]
\centering
	\includegraphics[width=\wide]{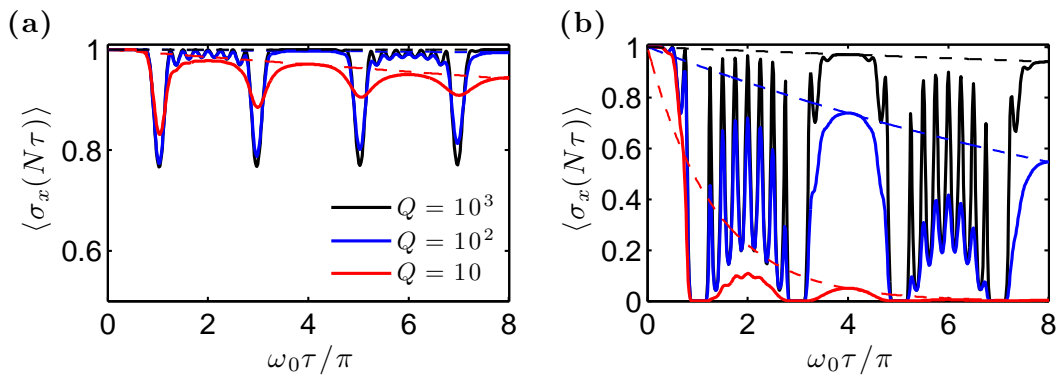}
\caption{Spin coherence 
for CPMG sequence with $N = 8$,
with undriven thermal oscillator at  temperature
$T = 10~\omega_0$ {\bf (a)}
and $T = 1000~\omega_0$ {\bf (b)}
and values of $Q$ shown.
Solid lines show full spin coherence with 
collapses and revivals, and
dashed lines show oscillator-induced
dephasing resulting in 
envelope decay
given by \eq{eq:Gphi}.
Here we took $\lambda/\omega_0 = 0.01$
and neglected intrinsic spin decoherence,
$T_1 = T_2 \rightarrow \infty$.
}
\label{fig:thermal}
\end{figure}
% ============================================

We plot the spin coherence due to thermal motion in the 
classical limit
$T \gg \omega_0$ in \fig{fig:thermal}.
The impact of the oscillator 
is greatest when
the pulse sequence is synchronized with
the cantilever frequency, $\tau = (2 k + 1)\pi/\omega_0$
with $k$ an integer.
At times $\tau = 2k\pi/\omega_0$, 
the accumulated phase
due the oscillator cancels within each
free precession time, so that
the accumulated phase variance averages
nearly to zero and the coherence revives.
We stress that this structure of collapse and revival
can arise from purely classical motion;
it is simply a consequence of averaging the
phase variance accumulated by the spin over
Gaussian distributed magnetic field fluctuations
with a characteristic frequency.
In addition to
collapses and revivals, the finite $Q$
of the cantilever also causes
dephasing of the spin which leads to an
exponential decay factor of the
envelope as $e^{-\Gamma_\phi \tau}$.
In the limit $Q \gg 1$ and $T> \omega_0$, the dephasing rate is given by 
\begin{equation}
\label{eq:Gphi}
	\Gamma_\phi \simeq
	3 N \eta^2
	 \gamma \left( \nth + \half \right),
\end{equation}
where $\eta = \lambda / \omega_0$
is the dimensionless coupling strength
and $\bar n_{\rm th} = (e^{\omega_0 / T} - 1 )^{-1}$
is the  thermal occupation number of the oscillator.
We provide
a derivation of \eq{eq:Gphi}  in \ref{app:highQ}.
Increasing $Q$ not only
increases the depth of the collapses
in spin coherence due to the oscillator, 
but also decreases the overall spin dephasing
resulting in more complete revivals, 
as shown in \fig{fig:thermal}.
We also see that increasing the temperature 
increases both the depth of collapse
and the dephasing.
Below in \secref{sec:sensitivity} we use these
results to calculate the lowest temperature motion
that can be detected, characterized by the 
phonon number
sensitivity at the optimal pulse timing $\tau = \pi /\omega_0$.

\subsection{Driven motion}
\label{sec:driven}

\begin{figure}[htb]
\begin{centering}
	\includegraphics[width=\figwidth]{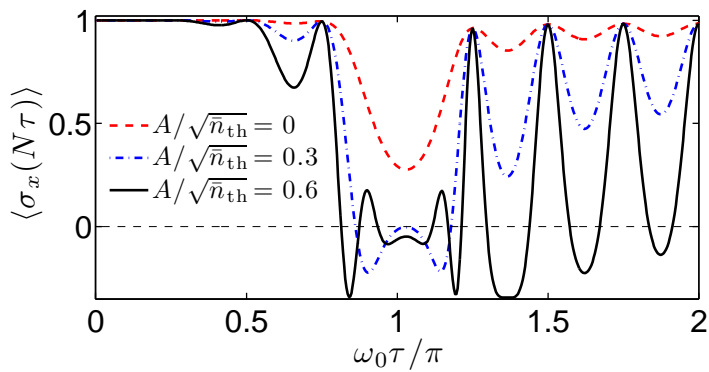}
\caption{Spin coherence from combined thermal and driven motion
for drive amplitudes shown.
For a weak drive, both driven and thermal contributions
are important.
The dips in the spin coherence below zero
arise from driven motion, described by
the Bessel function in \eq{eq:sxBessel}.
Parameters are $\omega_0/2\pi = 1$ MHz, 
$T = 50\ \omega_0$, $Q = 100$.
}
\label{fig:driven}
\end{centering}
\end{figure}

It is straightforward to include
the effects of
a classical drive through $H_{\rm dr}$
in \eq{eq:H}.
This simply adds
a classical deterministic contribution to $\hat X(t)$,
and we can decompose the
accumulated phase in
\eq{eq:phase}  it as
$\hat \phi =  \phi_{\rm dr} +  \hat \phi_{\rm th} $
where 
\begin{equation}
	\phi_{\rm dr} = \lambda A \int dt  \cos{( \omega_0 t + \theta_0 )} f(t,\tau)
\end{equation}
is the classical accumulated 
phase due to the drive.  
Here, $A$ is the 
dimensionless amplitude of driven motion
and $\theta_0$ is its  phase at
the start of a particular measurement.
We assume that the cantilever
drive is not phase-locked to the 
pulse sequence, so $\theta_0$
is random and
uniformly distributed between 0 and $2\pi$.
Using \eq{eq:D}
and averaging over $\theta_0$ we obtain
\begin{equation}
\label{eq:sxBessel}
	\avg{\sx(t=N\tau)} = J_0[a(\tau)] e^{-\chi_N(\tau)}
\end{equation}
where $J_0$ is the zeroth order 
Bessel function \cite{Lange:2011jo},
$a(\tau) = \eta A  \sqrt{2 F(\omega_0 \tau)}$,
and $\chi_N(\tau)$ is the thermal contribution given by
\eq{eq:chi}.
For a strong drive, thermal fluctuations
are unimportant and the 
signal is given by the
Bessel function.
For a weak drive,
comparable to thermal motion with
$\abs{A}^2 \sim \nth$,
both thermal and driven contributions
may be important as illustrated in 
\fig{fig:driven} and observed 
in experiment \cite{Kolkowitz:2012iw}.
In \fig{fig:driven} we see that, 
unlike thermal motion (see \fig{fig:thermal}),
driven motion can lead to dips in the spin coherence
below zero.
In the remainder of the paper we focus on detecting
undriven thermal or quantum motion with
the drive switched off.
% ======================== figure =================
% ============================================

\section{Phonon number sensitivity}
\label{sec:sensitivity}

In this section we discuss the sensitivity limits
of the spin used as a detector of undriven
mechanical motion.
By comparing the signal 
from thermal motion to
the relevant noise sources,
we obtain the phonon number sensitivity.
We then discuss the sensitivity in
several limits relevant to experiments.

\subsection{Signal}

The impact of an undriven thermal oscillator
on the spin coherence in a 
spin echo or CPMG measurement sequence
is described by \twoEq{eq:S}{eq:Dtherm}.
In addition to its coupling to the oscillator, the spin
is also coupled to an environment which 
leads to intrinsic decoherence and degrades
the signal.
For an NV center,
decoherence or $T_2$ processes are caused by
a 1\% natural abundance of
$^{13}$C nuclear spins in 
the otherwise $^{12}$C lattice.
Flip-flop processes between
pairs of these nuclear spins produce low frequency
magnetic noise
which leads to decoherence
of the form $e^{-N (\tau/T_2)^3}$ 
for a CPMG  sequence with $N$ pulses 
\cite{Taylor:2008cp,deSousa:2009fp}.
Note that $T_2$ here refers to the decoherence time 
in a spin echo sequence (i.e.~$N$=1),
typically $\sim$ 100 $\mu$s in natural diamond
and up to $\sim$ 2 ms in isotopically pure diamond 
\cite{Balasubramanian:2009jz}.
An added benefit of multipulse sequences
is the enhanced spin coherence time,
$\tilde T_2 = N^{2/3} T_2$, due to dynamical decoupling 
\cite{deLange:2010ga}.
Finally, spin-lattice relaxation due to
phonon processes leads to 
exponential decay on a timescale $T_1$,
typically $\sim$ 1 ms at room temperature
and up to $\sim 200$ s at 10 K \cite{Jarmola:2011wf}.
Including these intrinsic sources of spin
decoherence, as well as the oscillator-induced
decoherence $\Gphi$ given in \eq{eq:Gphi},
the probability
to find the spin in its initial state
given in \eq{eq:S}
is modified as
\begin{equation}
\label{eq:Sdetect}
	P(t=N \tau) = \frac{1}{2} \left(
	1 + e^{-N \left( \tau/T_1 + (\tau/T_2)^3\right) }
	e^{-\Gphi \tau} \right)
	- {\mathcal S}(\tau),
\end{equation}
where we have isolated the
coherent signal due to the oscillator,
\begin{equation}
\label{eq:signal}
	{\mathcal S}(\tau) = 
	\half e^{-N \left( \tau/T_1 + (\tau/T_2)^3\right)}
	e^{-\Gphi \tau}
	\left( 1 - e^{-\left( \chi_N(\tau) - \Gphi \tau \right)} \right).
\end{equation}
Note that we have accounted for the oscillator-induced
decoherence $\Gphi \tau$ which
diminishes the coherent signal
we are interested in.

We can obtain a simple analytic expression
for the signal
in the limit $Q \gg 1$.
In this limit the oscillator
spectrum is well-approximated
by Lorentzians at $\omega = \pm \omega_0$,
\begin{equation}
\label{eq:SXapprox}
	\bar S_X(\omega) \simeq 
	\frac{\gamma  \left( \bar n_{\rm th} + 1/2\right)}
	{(\omega - \omega_0)^2 + \gamma^2/4}
	+
	\frac{\gamma  \left( \bar n_{\rm th} + 1/2\right)}
	{(\omega + \omega_0)^2 + \gamma^2/4}.
\end{equation}
Using \eq{eq:SXapprox} with \eq{eq:chi} we obtain
a compact analytic
expression for $\chi$ with no further approximation,
which we provide in \ref{app:highQ}.
We choose the pulse timing $\tau$ to maximize the
impact of the oscillator motion on the spin coherence,
providing optimal sensitivity.
This is achieved by 
setting $\tau = \pi/\omega_0$,
flipping the spin every half period of
the oscillator and resulting in the
maximum accumulated phase variance.
For $N \gg1$,
the
filter function with $\tau = \pi/\omega_0$
is well-approximated by a Lorentzian centered at $\omega_0$
of bandwidth $b \omega_0 / N$,
where $b \simeq 1.27$.
Together with \eq{eq:SXapprox} this yields
\begin{equation}
\label{eq:chiLargeN}
	\chi_N(\pi/\omega_0) \simeq \frac{16 \eta^2 Q N}
	{\pi\left( 1 + b Q / N \right)} \left( \nth + \half \right),
\end{equation}
and substituting this into \eq{eq:signal} we obtain the signal.

\subsection{Sensitivity}

To find the sensitivity we must
account for noise.
We combine spin projection and
photon shot noise into a single
parameter $K$
so that the
noise averaged over $M$
measurements is 
$\sigma = 1 / K\sqrt{M}$,
where $M = \ttot / N\tau$
is the
number of measurements
of duration $N \tau$
that can be performed in a total time
$\ttot$.
It follows that the minimum 
number of phonons that we can 
resolve in a given time $t_{\rm tot}$ is 
\begin{equation}
	\nimp = \frac{\sigma}{\abs{d {\mathcal S} / d \bar n_{\rm th}}},
\end{equation}
and the corresponding phonon number sensitivity
is
$\sens = \nimp \sqrt{\ttot}$.
Using \twoEq{eq:signal}{eq:chiLargeN} with
$\tau = \pi/\omega_0$ we obtain
\begin{equation}
\label{eq:sens}
	\sens \simeq
	\frac{ \pi^{3/2} }{8K \eta^2 Q N}
	e^{ N / N_{\phi}}
	 \left( 1 + \frac{b Q}{N} \right)
	\frac{1}{\sqrt{\omega_0/N}},
\end{equation}
where  we have expressed the total spin dephasing 
in terms of a single pulse number,
\begin{equation}\label{eq:Nphi}
	N_{\phi}  = \left[
	\frac{\pi }{\omega_0 T_1 } + 
	 \left(\frac{ \pi}{ \omega_0 T_2} \right)^3
	+ \frac{3\pi \eta^2}{Q}  \left(\nth + \half \right)
	\right]^{-1},
\end{equation}
which combines both intrinsic 
and oscillator-induced decoherence.
\twoEq{eq:sens}{eq:Nphi} reflect the competition between
the oscillator damping rate $\gamma = \omega_0/Q$,
the intrinsic decoherence times $T_1$ and $T_2$ of the spin,
and the measurement bandwidth $b \omega_0 / N$.
It is clear from \eq{eq:chiLargeN} that
increasing the number of pulses increases the
coherent signal due to the oscillator;
however, this also leads to increased
spin decoherence.
As a result, the resolvable phonon number is
minimized at an optimal number of pulses,
\begin{equation}
\label{eq:Nopt}
	\Nopt = N_{\phi} - b Q + \sqrt{N_\phi^2 + 6 b Q N_{\phi} + 
	\left( b Q \right)^2}.
\end{equation}
Note that the optimal pulse number is always
set by the spin decoherence, $\Nopt \sim N_{\phi}$, with
only a prefactor of order one depending on $Q$.
Neglecting pulse imperfections,
the optimized sensitivity is determined by an
interplay of $Q$, $T_1$ and $T_2$ in \eq{eq:sens}.
In practice, the optimal pulse number may be
very large due to long spin coherence times, and
pulse errors may play a role as discussed further below.

% ======================== figure =================
\begin{figure}[htb]
\centering
	\includegraphics[width=\wide]{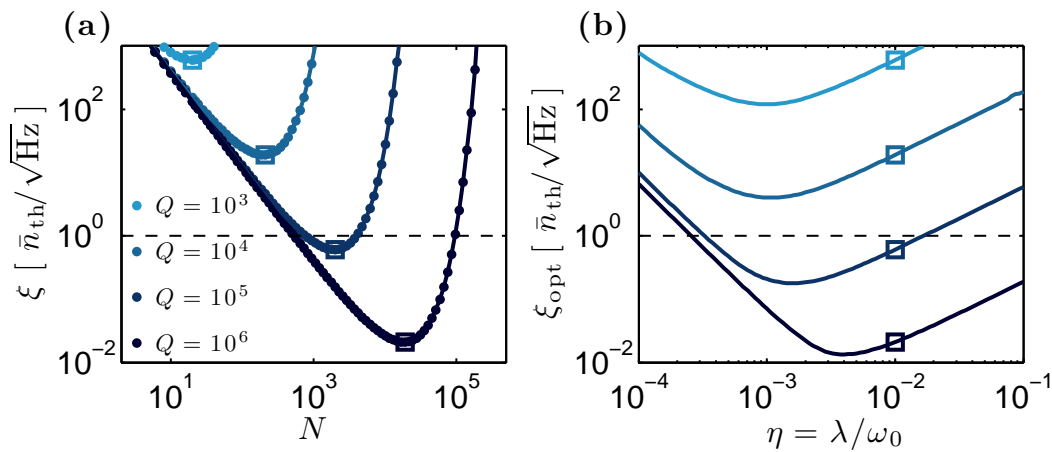}
\caption{{\bf (a)} Phonon number sensitivity $\xi$ 
versus the number of pulses $N$ for values
of $Q$ shown and $\lambda/\omega_0 =  0.01$.
Lines show the analytic result in \eq{eq:sens} and
points show the full numerical result using
\twoEq{eq:chi}{eq:SX}.
Squares mark the sensitivity at the optimal pulse
number $\Nopt$.
{\bf (b)} Sensitivity optimized with respect
to $N$, versus 
the coupling strength $\lambda/\omega_0$ for the
same values of $Q$ as in {\bf (a)}.
Squares mark the optimized sensitivity at $\lambda/\omega_0 = 0.01$,
corresponding to the squares in {\bf (a)}.
The dotted lines mark a sensitivity of $\xi = 1/\sqrt{\rm Hz}$.
Parameters in both plots are $\omega_0/2\pi = 1$ MHz,
$T_2 = 100\ \mu$s, $T_1 = 100$ ms, $T = 4$ K and $K = 0.3$.
}
\label{fig:sens}
\end{figure}
% ============================================

In \fig{fig:sens} we plot the sensitivity as a 
function of pulse number $N$,
and the optimized sensitivity as a function of
coupling strength $\lambda$.
To check the validity of the above approximations it is
straightforward to 
calculate the phonon number
sensitivity directly 
from \twoEq{eq:chi}{eq:SX}.
The numerically exact sensitivity is shown
in \fig{fig:sens} in agreement
with our analytic results. 
In the remainder of this section we 
discuss the sensitivity in
several experimentally relevant limits.

\subsection{Optimal sensitivity and cooperativity}

An important limit for current experiments 
is one where the spin coherence is
much longer than the oscillator coherence during the
measurement, corresponding to
$N_\phi>Q$. We assume that the spin coherence is dominated by
intrinsic sources described by
$T_1$ and $T_2$, and that
the oscillator-induced
spin decoherence $\Gamma_\phi$ can be neglected,
well-justified in the limit of weak coupling.
Within these limits,  
the optimal number of pulses
is $\Nopt \sim N_\phi$
and the optimized sensitivity is
\begin{equation}
\label{eq:sensCoop}
	\sens_{\rm opt} \simeq \frac{\pi^{3/2}}{8K C \sqrt{\omega_0 / N_{\phi}}},
\end{equation}
where the cooperativity is
\begin{equation}
\label{eq:C}
	C = \frac{\lambda^2 \tilde T_2}{\gamma},
\end{equation}
and $\tilde T_2 = \Nopt^{2/3} T_2$ is the 
enhanced spin coherence time due to decoupling.
For a large number of pulses, the enhanced 
spin coherence $N^{2/3}T_2$ may be very long,
and ultimately
the spin coherence may be limited by $T_1$ which
is not suppressed by decoupling.
In this case 
\twoEq{eq:sensCoop}{eq:C} are simply 
modified by $\tilde T_2 \rightarrow T_1$. 
The cooperativity parameter $C$ is ubiquitous in quantum
optics, and marks the onset of Purcell enhancement
in cavity quantum electrodynamics. 
In the present case, $C > 1$ is the requirement
for a single phonon
to strongly influence the spin coherence, leading
to a measurable signal despite the relatively short coherence
time of the oscillator.
The condition $C>1$ to resolve a single phonon
can be simply understood:
if the spin coherence is much longer
than the oscillator coherence, i.e.~$Q \ll N_\phi$, the accumulated phase
variance increases at a rate $\sim \lambda^2 / \gamma$
(see \eq{eq:chiLargeN} with sequence time $N \tau \sim N/\omega_0$)
and the maximum interrogation time (assuming
that oscillator-induced decoherence is negligible) 
is $\tilde T_2$.
 
With feasible experimental parameters,
$\tilde T_2\sim T_1 \sim 10$ ms,
$\lambda/2\pi \sim 150$ Hz,
$\omega_0/2\pi \sim 1$ MHz
and $Q \sim 1000$, 
a cooperativity of 
$C \sim 1$ can be reached.
In current experiments, NV centers
exhibit a 30\% contrast in
spin-dependent fluorescence,
and collection efficiencies
of 5\% are realistic \cite{Taylor:2008cp,Robledo:2011fs}.
These parameters yield $K \sim 0.3$ and an optimal
phonon number sensitivity of
$\sens_{\rm opt} \sim 1 / \sqrt{\rm Hz}$ 
with  $N \sim N_\phi \sim 15000$ pulses.
Due to long spin coherence times $T_1$ and
$T_2$,
the optimal pulse number $N_\phi$
may be very large, and in practice 
finite pulse errors may play an important role
in limiting the spin coherence.
For example if the number of pulses is limited to $N \sim 1000$,
a sensitivity of $\xi \sim 3 / \sqrt{\rm Hz}$ 
can be reached.
We discuss this further below when we
calculate the signal due to zero point motion.

\subsection{Ideal oscillators and ideal spin qubits}
\label{sec:idealLimits}

While the cooperativity regime describes
an important
part of parameter space, it is useful to briefly consider
two more simple limits that describe features
in \fig{fig:sens}.
First, we consider a harmonic oscillator
that remains coherent for a much longer time than
the entire pulse sequence, satisfying
$Q \gg N$.
In this limit, the long oscillator coherence time
plays plays no role and
the optimal sensitivity is limited only 
by the spin coherence,
$\sens_{\rm opt} \sim 1 / (K \lambda^2 \tilde T_2^2 \sqrt{\omega_0 / N})$.
This limit can be seen on the left side of \fig{fig:sens}a,
where the sensitivities
for different values of $Q$ fall on the same curve
at low pulse numbers $N$.

Finally, we consider the limit
of very strong but incoherent coupling where
the spin decoherence is dominated by 
the oscillator, i.e. $\Gphi$ becomes larger than
$1/T_1$ and $1/T_2$.
This limit is reached when either 
the intrinsic spin decoherence is negligible 
or for very strong coupling,
$\eta^2 \nth \gg Q /  (\omega_0 T_2)^3, Q / \omega_0 T_1$.
In this limit, the coherent
signal is large due to strong coupling, but
saturates at a low number of pulses;
further increasing the
coupling strength only increases the
oscillator-induced decoherence,
reducing the signal. 
This is reflected in
\fig{fig:sens}b, where we see that increasing 
the coupling strength larger than $\eta^2 > 1 / \gamma \nth T_1$ 
no longer improves the optimized sensitivity
but instead degrades it.

\section{Detecting quantum motion}
\label{sec:zpm}

Above we found that for realistic experimental
parameters, 
a single phonon can be resolved 
in one second of averaging time.
This raises the intriguing question of whether a single spin
can be used to sense the quantum zero point motion
of an oscillator in its ground state. 
It also implies that we must consider the effect of 
measurement backaction, which we have so far ignored in our discussion.
To address these questions we analyze 
the experimentally relevant scenario
where the spin is used to detect the
motion of a mechanical resonator
which is
externally cooled close to its ground state.

\subsection{Measuring a cooled oscillator}

Even at cryogenic temperatures, a mechanical oscillator 
of frequency $\omega_0 / 2\pi \sim$ MHz
has an equilibrium occupation number $\nth$
much larger than one.  
For this reason we assume that
the mechanical oscillator is cooled
from its equilibrium occupation $\nth$
to a much lower value $\bar n_0\sim 1$ using either
optical cooling techniques \cite{Marquardt:2007dn}
or the driven spin itself \cite{Rabl:2009fz,Rabl:2010cm}.
An important consequence of
cooling below the environmental
temperature is the effective reduction in
$Q$ of the oscillator.
For an oscillator coupled to both a thermal environment
and an external, effective zero temperature source for cooling,
the mean phonon number satisfies
\begin{equation}
     \avg{\dot n} = - (\gamma + \gamma_{\rm cool}) \avg{n} + \gamma \nth,
\end{equation}
where $\gamma_{\rm cool}$ is the cooling rate.
The steady state occupation number is
\begin{equation}
    \bar n_0= \avg{n}(t\rightarrow \infty) = \frac{ \gamma \nth}{\gamma +\gamma_{\rm cool}},
\end{equation}
and in order to maintain $\bar n_0 < 1$
we require $\gamma_{\rm cool} > \gamma\nth$.
As a result,
the relevant decoherence rate of the oscillator
is the rethermalization rate $\gamma \nth$.
For this reason,
to calculate the signal from
a cooled oscillator we replace 
the equilibrium thermal occupation number $\nth$ by the effective
occupation $\bar n_0 \rightarrow 0$ in all expressions,
while at the same time replacing the intrinsic $Q$ by the reduced,
effective
quality factor $\Qeff = \omega_0/\gamma_{\rm cool}\approx Q  / \nth$.

\subsection{Single shot readout}

In \secref{sec:sensitivity} we calculated the sensitivity $\xi$, 
which reflects the minimum detectable phonon number $n_{\rm min}$ 
that can be resolved in one second of averaging time.
For the following discussion it is useful to convert the sensitivity to a
minimum detectable phonon number per single measurement shot,
$\nssr = \sens / \sqrt{(N\pi/\omega_0)}$,
where we have taken the total measurement time to be $\ttot = N\tau$ and
$\tau = \pi/\omega_0$.
Assuming single shot spin readout ($K \rightarrow 1$),
which has been demonstrated at low temperature \cite{Robledo:2011fs},
and using \eq{eq:sens} we obtain
\begin{equation}
\label{eq:SingleShotSensitivity}
    \nssr = \frac{\pi e^{{N/N_{\phi}}}}{8 \eta^2N\Qeff} 
    \left( 1 + \frac{b\Qeff}{N}\right) 
    \sim \frac{1}{C_{\rm eff}},
\end{equation}
where $C_{\rm eff} = \lambda^2 \tilde T_2 / \gamma \nth$ is the 
reduced, effective cooperativity. 
We see that under the assumption $N\sim \Nopt \gg \Qeff$, 
the ability to resolve ground state fluctuations of a cooled oscillator within a few spin measurements
requires $C_{\rm eff}>1$, which is the same
strong cooperativity condition required to perform a 
quantum gate between two spins 
mediated by a mechanical oscillator \cite{Rabl:2010gza}. 
Alternatively, $\nssr$ 
corresponds to the occupation number required to produce a
signal ${\mathcal S}$ of order one in \eq{eq:signal}.
It provides a convenient way to directly
compare the sensitivity with the backaction due
to sequential measurements, as discussed below.
% ======================== figure =================
\begin{figure}[htb]
\begin{centering}
	    \includegraphics[width=\wide]{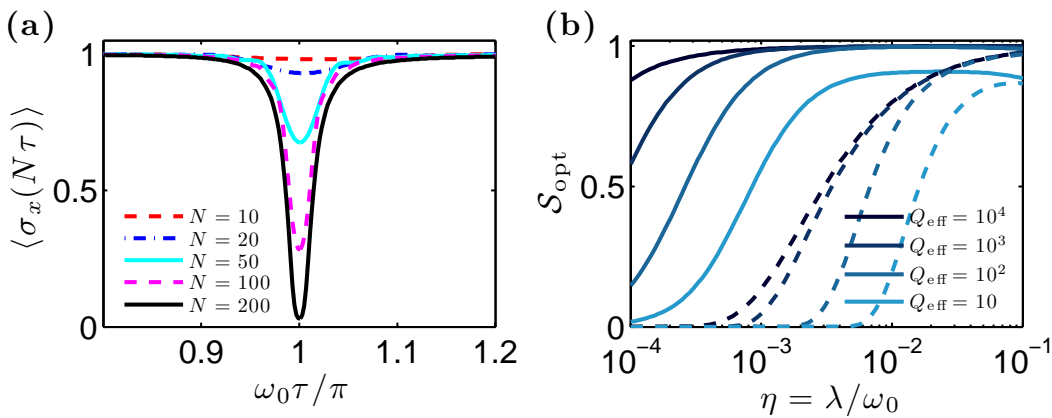}
\caption{{\bf (a)} Spin coherence with $\bar n_0 \sim 0$ for increasing
pulse number and $\Qeff = 100$ with $\lambda/\omega = 0.01$.
{\bf (b)} Optimal signal as defined in \eq{eq:signal} 
from zero point motion. 
Solid lines show optimal signal assuming
unlimited pulse number, while dashed lines
include a simple treatment of pulse errors with
$N_c = 1000$ as described in the text.
Parameters are $T_2 = 100$ $\mu$s, $T_1 = 100$ ms,
$\omega_0/2\pi = 1$ MHz.
}
\label{fig:zpm}
\end{centering}
\end{figure}
% ============================================

In \fig{fig:zpm} we plot the calculated signal 
due to zero point motion, assuming that the mechanical
oscillator is cooled near its ground state
$\bar n_0=0$ and using the reduced quality factor $\Qeff$. 
These plots show that the intrinsic coherence times
typical for
NV centers are more than sufficient to resolve single
phonons provided enough pulses can be applied to
exploit the full spin coherence.
In practice, the limiting factor is likely to be finite pulse
errors, which limit the absolute number of pulses that
can be applied before losing the spin coherence.
To estimate the effect of finite pulse errors,
we include the calculated signal assuming additional
spin decoherence of the form $e^{-N/N_c}$
with a cutoff pulse number $N_c$.
Pulse numbers of $N \sim 160$ have been
demonstrated in experiment 
\cite{deLange:2010ga}, and with further
improvements this can be increased
to more than $N \sim 1000$.
Based on this we plot the modified signal using
$N_c \sim 1000$ and find that even with a
limited number of pulses,
zero point motion results in a significant signal for 
realistic coupling strengths.

\subsection{Backaction}
\label{sec:ba}

The result that a single spin magnetometer can resolve
the quantum zero point motion of a mechanical oscillator
calls for a discussion of measurement backaction.
We begin by noting that,
despite the linear coupling of the spin to the oscillator
position in \eq{eq:H}, 
the described measurement protocol is sensitive to the \emph{variance}
of the accumulated phase $\sim \langle \hat X^2\rangle$,
which we obtain by averaging independent spin measurements.
As a result, our approach does not correspond
to standard continuous position measurement \cite{Caves:1980jp},
nor
does it implement a quantum nondemolition measurement of the 
phonon number,
since the interaction in \eq{eq:H}
does not commute with $\hat n$.
In principle, by cooling between measurements our approach
may be used to measure the phonon number with arbitrary precision.
Nonetheless, the effect of the spin's backaction on the oscillator
is both a practical issue and interesting in itself,
and could be used to prepare
nonclassical mechanical states.
We describe two possible
approaches to observe the influence of measurement backaction on the oscillator.

First, we consider directly probing the projective nature of
the measurement.
For simplicity we assume
that the oscillator is initially in its ground state
and decoupled from the environment,
and assume single shot spin readout.
In a single measurement sequence,
the oscillator
experiences  a spin-dependent
force according to \eq{eq:Hint}. 
Measuring $\avg{\sx} = \pm 1$ at the end of the sequence
projects the oscillator onto a superposition of coherent
states \cite{Steinke:2011ig,Tian:2005cu},
\begin{equation}\label{eq:State1}
	\ket{\psi_\pm}=  \frac{ \ket{i\alpha} \pm \ket{-i\alpha} }
	{\sqrt{2\pm 2 e^{-2\alpha^2}}},
\end{equation}
where  $\alpha = N \lambda / 2\omega_0$ is the total 
displacement for a sequence
of $N \gg 1$ pulses and $\tau = \pi/\omega_0$. 
The probabilities to measure $|\pm\rangle$ are given by 
\begin{equation}
	p_\pm =\frac{1}{2}\left( 1\pm e^{-2\alpha^2}\right),
\end{equation}
which shows, consistent with
the discussion above, that for a measurement strength $\alpha>1$  
the oscillator in its ground state
can significantly affect the spin dynamics.
To observe the backaction of this measurement on the
oscillator we can perform a second spin measurement,
which is sensitive to  
the state of the oscillator conditioned on the first measurement.
In principle, by using techniques developed in 
cavity quantum electrodynamics, this procedure
can be used to fully reconstruct the
conditionally prepared oscillator state \cite{Deleglise:2008gt}.

Let us now consider an alternative, 
indirect way to observe backaction 
by performing many successive measurements.
Again beginning with the oscillator near its ground state,
the first measurement  projects the oscillator into one of the states $|\psi_\pm\rangle$.
By averaging over the two possible spin measurement outcomes, 
the resulting mixed oscillator state is 
\begin{equation}
	\rho_{\rm osc} =  p_+ \ket{\psi_+}\bra{\psi_+}+p_- \ket{\psi_-}\bra{\psi_-},
\end{equation}
and we see that on average the oscillator energy has increased by $|\alpha|^2$. 
Repeating this measurement many times,
without cooling between measurements,
the oscillator amplitude undergoes a 
random walk of stepsize $\pm \alpha$, and
on average the phonon number 
increases approximately linearly in time.
This corresponds to backaction heating described
by an effective diffusion rate,
\begin{equation}
    	D_{\rm ba} = \frac{N \eta^2\omega_0}{4\pi}.
\end{equation}
% ======================== figure =================
\begin{figure}[tb]
\begin{centering}
	\includegraphics[width=\figwidth]{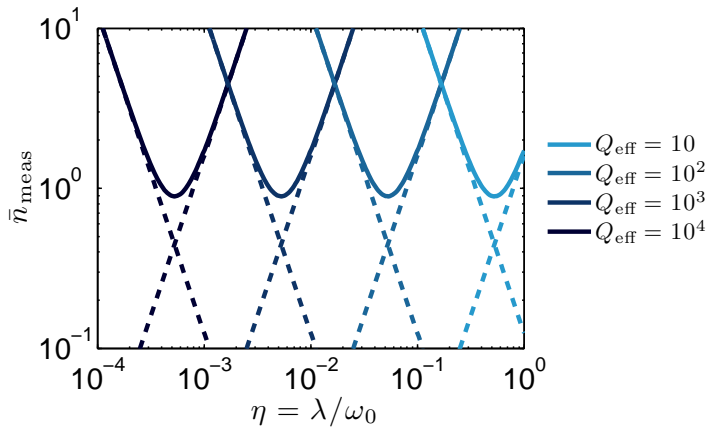}
\caption{
Solid lines show
total inferred phonon number
given by \eq{eq:nadd} from
combined phonon resolution and backaction heating.
Dashed lines show sensitivity and heating contributions.
For each value of $\Qeff$ we set $N = \Qeff / 5$.
}
\label{fig:backaction}
\end{centering}
\end{figure}
% ===========================================
Combining the measurement backaction with intrinsic 
mechanical dissipation and external cooling,
the average occupation number
satisfies
\begin{equation}
     \avg{\dot n} = - (\gamma + \gamma_{\rm cool}) \avg{n} + \gamma \nth + D_{\rm ba},
\end{equation}
and for $\gamma_{\rm cool}\gg\gamma$ the steady state phonon number added due to backaction
is
\begin{equation}
\label{eq:nadd}
    \nadd = \frac{D_{\rm ba}}{\gamma_{\rm cool}} = \frac{N \Qeff \eta^2}{4\pi}.
\end{equation}
We see that increasing the coupling strength not only improves
the single shot resolution $\nssr$, but also
leads to backaction heating of the oscillator.
For sufficiently strong coupling, the
steady state backaction phonon number
$\nadd$ exceeds the phonon number resolution,
and the inferred phonon number is determined by backaction.
We thus take the sum  
$\nmeas = \nssr + \nadd$ as a measure of
the minimum inferred phonon number.
Note that for simplicity in this discussion we have assumed
the limit $N\ll Q_{\rm eff}$, in which the oscillator is coherent
within each measurement sequence.
Within this limit we find 
\begin{equation}
	\nmeas = \frac{\pi \alpha}{8 \eta^2N^2} +\frac{N \Qeff \eta^2}{4\pi}.  
\end{equation}
The total inferred phonon number $\nmeas$ is shown
in \fig{fig:backaction} as a function of the coupling 
parameter $\eta$ and a fixed number of pulses $N= Q_{\rm eff}/5$. In this case  
$\nmeas$ is  minimized for $\eta\sim 1/\sqrt{Q_{\rm eff}}$, 
where it reaches a value of $\nmeas\sim\mathcal{O}(1)$.
Observing this minimum in the 
phonon number resolution as a function
of coupling strength
would provide an indirect signature
of measurement backaction.
This observation 
may be more feasible in near term experiments
than directly observing projective
backaction as discussed above.

\section{Summary and conclusions}

We have presented the sensitivity limits of a novel position
sensor consisting of a single spin.
For realistic experimental parameters, we predict that a single
NV center in diamond can be used to resolve single phonons
in a cooled, magnetized mechanical cantilever.
The condition to resolve single phonons 
is that of strong effective cooperativity, 
the same condition needed
to perform a quantum gate between two spins 
mediated by a mechanical
oscillator.
For even stronger coupling, the backaction
of the spin on the oscillator can be probed directly or
indirectly, and
used to prepare nonclassical mechanical states.

\ack

This work is supported by NSF, CUA, DARPA and the Packard Foundation.
SDB acknowledges support from 
NSERC of Canada and ITAMP.
SK acknowledges support by the DoD through the NDSEG Program, 
and the NSF through the NSFGRP under Grant No. DGE-1144152. 
QPU acknowledges support from Deutschen Forschungsgemeinschaft.
PR acknowledges support by the Austrian Science Fund (FWF)
through SFB FOQUS and the START grant Y 591-N16.

\appendix

\section{Analytic signal for thermal motion in high $Q$ limit}
\label{app:highQ}

Here we sketch the derivation
of \twoEq{eq:Gphi}{eq:chiLargeN}. 
The impact of the oscillator on the spin coherence is
given by \twoEq{eq:chi}{eq:SX}, 
\begin{equation}
\label{eq:chiA}
	\chi_N(\tau) =
	2\omega_0 \gamma \lambda^2 \int \frac{d\omega}{2\pi}
	\frac{	F(\omega \tau)}{\omega^2}
	 \frac{  \omega \coth(\omega/2T) }
	{(\omega^2-\omega_0^2)^2 + \gamma^2 \omega^2}.
\end{equation}
To perform this integral it is useful to decompose the
filter function as
\begin{equation}
\label{eq:filterRewrite}
	 F(\omega\tau) = 1 - \cos(N\omega\tau)
	+ \sum_{j=0}^{N-1} (-1)^j \bigg[
	\left( 1-\cos(\omega s_j)\right)  - j \left( 1 - \cos(\omega t_j)\right)
	\bigg],
\end{equation}
where $s_j = (j + 1/2)\tau$ and $t_j = (N-j)\tau$.
We first consider the high temperature limit, $T \gg \omega_0$, in which
we can approximate $\coth(\omega/2T)\approx 2T/\omega$.
The result is a sum of integrals of the form
\begin{equation}
	4 T \omega_0 \gamma \lambda^2 \int \frac{d\omega}{2\pi}
	\frac{1-\cos(\omega t)}
	{\omega^2\left[ (\omega^2 - \omega_0^2)^2 + \gamma^2 \omega^2 \right]} 
	= \eta^2 (2\nth) q(t),
\end{equation}	
which can be done exactly.  
In the limit $Q \gg 1$ we obtain
\begin{equation}
	q(t) =
		\gamma t
		+ 
		\left( 1 - e^{-\gamma t/2} \cos(\omega_0 t) \right)
		- \frac{4 \gamma}{3 \omega_0}
		e^{-\gamma t/2} \sin(\omega_0 t).
\end{equation}
To calculate $\Gphi$ we need the spin coherence
at the revivals, given by $\chi_N(\tau = 4\pi/\omega_0)$.
To first order in $\gamma t$, we have 
$q(t = 4\pi  / \omega_0) \simeq 3 \gamma t / 2$,
and to this order the only nonzero term in \eq{eq:chiA}
is due to the $1 - \cos(N \omega\tau)$
terms in \eq{eq:filterRewrite}.  The result is \eq{eq:Gphi}.

Next we derive \eq{eq:chiLargeN} at $\tau = \pi/\omega_0$
in the limit $N \gg 1$.
Here we use the fact that the filter function
near $\tau \simeq \pi/\omega_0$ may be rewritten
for $N \gg 1$ as
$F(\omega\pi/\omega_0) \simeq 2 N^2 {\rm sinc}^2 \left[ \pi N (\omega-\omega_0)/2\right]$,
and in turn this function is well-approximated by its Lorentzian envelope,
\begin{equation}
\label{eq:Fapprox}
	F(\omega\pi/\omega_0) \simeq
	\frac{ (b\omega_0)^2/2}{ (\omega-\omega_0)^2 + (b\omega_0/N)^2/4},
\end{equation}
where we obtain the effective bandwidth $b\omega_0/N$ by fitting the
extrema of \eq{eq:Fapprox} to a Lorentzian which yields $b \simeq 1.27$.
At the collapse time $\tau = \pi/\omega_0$, we can approximate
$\bar S_X(\omega)$ by the Lorentzian spectrum given in \eq{eq:SXapprox}.
Using \twoEq{eq:SXapprox}{eq:Fapprox}, $\chi(\pi/\omega_0)$,
the integrand
is simply the product of two Lorentzians and 
performing the integration yields \eq{eq:chiLargeN}.

\section*{References}

\bibliography{NVcantilever1}
\bibliographystyle{iopart-num}

\end{document}